\documentclass[preprint,tightenlines,aps,floats,nofootinbib]{revtex4}  
  
\usepackage{epsf}
    
\preprint{  
\hbox to \hsize{  
\hfill$\vcenter{\hbox{\bf MADPH-04-1385}
	        \hbox{\bf OKHEP-04-02} 
	        \hbox{\bf UPR-1082-T} 
                \hbox{\bf hep-ph/0408120} 
                \hbox{August 2004}}$ } 
}  
  
\begin{document}  
  
\title{\vspace*{.75in}
Neutralino Relic Density in a Supersymmetric $U(1)'$ Model}
\author{
Vernon Barger$^1$, Chung Kao$^2$, Paul Langacker$^3$, and Hye-Sung Lee$^1$}
  
\affiliation{  
$^1$Department of Physics, University of Wisconsin,  
Madison, WI 53706 \\  
$^2$Department of Physics and Astronomy, University of Oklahoma,  
Norman,  OK 73019 \\
$^3$Department of Physics and Astronomy, University of Pennsylvania, 
Philadelphia, PA 19104
\vspace*{.5in}}
  
     
%
\thispagestyle{empty}  


\begin{abstract}  
 
We study properties of the lightest neutralino ($\chi^0$) and calculate its 
cosmological relic density in a supersymmetric $U(1)'$ model with a secluded $U(1)'$ breaking sector (the $S$-model).
The lightest neutralino mass is smaller than in the minimal 
supersymmetric standard model; for instance, $m_{\chi^0} \alt 100$ GeV in the limit that the $U(1)'$ gaugino mass is large compared to the electroweak scale.
We find that the $Z \chi^0 \chi^0$ coupling can be enhanced due to the singlino 
components in the extended neutralino sector.
Neutralino annihilation through the $Z$-resonance then reproduces the measured cold dark matter density over broad regions of the model parameter space.

\end{abstract}  
  
\maketitle

\newpage  

\section{Introduction}
  
Recent mapping of the Cosmic Microwave Background (CMB)     
anisotropy~\cite{WMAP} has provided precision information on the densities 
of matter and dark energy in the Universe.
The major part of the matter is non-relativistic and non-baryonic (cold and 
dark). 
When the Sloan Digital Sky Survey (SDSS) data on large scale structure are 
analyzed in combination with the Wilkinson Microwave Anisotropy Probe (WMAP) data, a cold dark matter (CDM) relic density 
\begin{eqnarray}
\Omega_{\rm CDM} h^2 = 0.12 \pm 0.01 \;\; {\rm (SDSS+WMAP)}
\label{eqn:sdsswmap}
\end{eqnarray}   
is found \cite{SDSS}.
This very restrictive range for $\Omega_{\rm CDM} h^2$ has significant impact 
on the allowed masses and couplings of cold dark matter particles. 

A neutral, stable, massive particle that interacts weakly is a natural 
candidate for CDM. The Standard Model (SM) does not have a particle with 
these properties but a supersymmetric model with $R$-parity conservation, 
one of the best motivated new physics possibilities, does.
In the Minimal Supersymmetric Standard Model (MSSM) the lightest neutralino 
is the favored lightest supersymmetric particle (LSP).

The lightest neutralinos existed abundantly in the early Universe in thermal  
equilibrium with other particles, where their pair annihilations were 
balanced by pair creation. As the Universe cooled, the neutralino density 
became Boltzmann-suppressed.   
Deviation from thermal equilibrium began when the temperature reached  
the freeze-out temperature  $T_f \simeq m_{\chi^0}/20$.  
After the temperature dropped to $\sim{1\over5}T_f$, the annihilation rate 
became equal to the expansion rate, and the neutralino relic density was 
$n_{\chi^0} = H/\left<\sigma v\right>$, where $H$ is the Hubble expansion 
rate at that temperature.  
Here $\left<\sigma v\right>$ is the thermally averaged cross section  
times neutralino velocity \cite{Jungman:pr.267.195}.   
The remaining CDM density relative to the critical density is
\begin{equation}  
\Omega_{\chi^0} = n_{\chi^0} m_{\chi^0} / \rho_{\rm c}  = H    
m_{\chi^0} / \left( \left< \sigma v \right> \rho_c \right). 
\end{equation}  

A viable CDM model must reproduce the recent precise measurement of $\Omega_{\rm CDM}$, 
preferably without fine-tuning of the model parameters. 
The full test of neutralino dark matter can be accomplished with the direct detection of the LSP 
in collider experiments and/or in elastic scattering experiments at 
underground detectors.
The parameters of the model must quantitatively explain $\Omega_{\chi^0} h^2 \simeq 0.12$ and the direct detection rates.

There have been extensive theoretical studies of the relic density 
\cite{relicdensity,Baer&Brhlik,Roszkowski}.
In the minimal supergravity model (mSUGRA) annihilations of stable 
supersymmetric particles can reproduce the right order of magnitude.
However, the allowed regions of mSUGRA parameter space are becoming 
increasingly constrained \cite{CMSSMconstraints} by the combination of the recent relic density and the LEP data and now 
allow only a few regions of parameter space including 
(i) pair annihilation of neutralinos with dominant bino composition through $A^0,H^0$-resonances
at high $\tan\beta$ (the so-called $A$-funnel) \cite{Barger&Kao},
(ii) neutralino-stau coannihilation \cite{sleptoncoannihilation}, and 
(iii) annihilation of neutralinos with mixed gaugino-Higgsino components in the focus point region \cite{focuspoint}.
A nearly pure bino LSP state can give the right size of the relic density 
\cite{purebino} and also satisfy the required radiative electroweak symmetry 
breaking (EWSB) in unified models \cite{unified}.
In $SO(10)$ grand unified supergravity models with Yukawa unification, 
it is more difficult to obtain the required relic density, but it is still possible with specific scalar mass patterns \cite{Auto:hep-ph/0407165}.

The MSSM suffers from the $\mu$-problem \cite{muproblem}
and the lack of a 
sufficiently strong first order phase transition for electroweak baryogenesis 
(EWBG) over most of the parameter space\footnote{To have a strong first-order 
phase transition in the MSSM, the light Higgs mass should be only slightly 
above the LEP experimental bound and the light stop should be lighter than the top \cite{stop}.}.
The missing ingredients of possible TeV-scale new physics that would cure 
these problems may modify the properties of the CDM candidate.
The Next-to-Minimal Supersymmetric extension of the Standard Model (NMSSM) 
\cite{NMSSM1} avoids the problems by introducing an additional SM singlet 
Higgs and a discrete symmetry ${\bf Z_3}$, but it then suffers from the 
cosmological domain wall problem \cite{domainwall}.

A natural extension of the MSSM that avoids the above difficulties is a 
supersymmetric $U(1)'$ model, with Higgs singlets to break the additional 
Abelian symmetry spontaneously at the TeV scale\footnote{For another 
extension of the MSSM using a discrete symmetry but free of domain wall 
problem, see the nMSSM model \cite{nMSSM2, Menon:hep-ph/0404184}.}.
Additional $U(1)$ gauge symmetries are predicted in many types of new 
theories, including superstrings \cite{Langacker:2004wf}, grand unification, 
extra dimensions \cite{Masip:1999mk}, dynamical symmetry breaking \cite{DSB}, 
and the little Higgs model \cite{LittleHiggs}.
In this letter we investigate the properties of the lightest neutralino 
and evaluate its relic density\footnote{For an earlier study of the relic 
density in a $U(1)'$ model with one singlet, in a different approach and 
framework, see Ref. \cite{deCarlos:plb.407.12}.} in an extended model of the 
MSSM with an extra $U(1)$ gauge symmetry and four extra Higgs singlets 
($S, S_1, S_2$, and $S_3$) \cite{Erler:prd.66.015002}.
The superpotential is\footnote{For a $U(1)'$ model with only one singlet, see Ref. \cite{Cvetic:prd.56.2861}. It is more difficult to obtain $M_{Z'} \gg M_Z$ in such models.}
\begin{equation}
W = h_s S H_1 H_2 +\lambda_s S_1 S_2 S_3,
\end{equation}
where $h_s$ and $\lambda_s$ are dimensionless parameters.
We call this the $S$-model.

The $S$-model solves the $\mu$-problem by generating an effective 
$\mu$ parameter
\begin{equation}
\mu_{\rm eff} = h_s \left<S\right>.
\end{equation}
The model is free from the domain wall problem since there is no discrete symmetry 
\cite{Suematsu:ijmpa.10.4521, Cvetic:prd.56.2861, Erler:prd.66.015002}.
The $S$-model can also provide a sufficiently strong first order phase transition 
and new sources of CP violation for EWBG \cite{Kang:hep-ph/0402086}.
The $Z'$ has a large mass\footnote{
The CDF limit is $M_{Z'} \agt 500 - 800$ GeV, depending on the model 
\cite{CDF,PDG}.} generated by singlet Higgs fields, $S_{1,2,3}$, which 
acquire large (TeV scale) VEVs for small $\lambda_s$ because of an almost $F$ 
and $D$ flat direction. 
These multiple singlets allow $\mu_{\rm eff}$ to be of the electroweak scale 
while keeping the $Z'$ heavier than the experimental limit.

Electroweak symmetry breaking is driven by electroweak scale trilinear terms.
This leads to $\tan\beta \equiv v_2/v_1 \simeq 1$, while solutions without 
unwanted global minima at $\left<H^0_i\right> = 0$ typically have  
$\left<S\right> \alt 1.5 \left<H^0_i\right>$ \cite{Erler:prd.66.015002}.
However, both of these conditions can be relaxed somewhat \cite{Han:hep-ph/0405244}.
All dimensionful supersymmetry-breaking parameters are close to the 
electroweak scale. The squark and slepton masses are similar to those of 
the MSSM, and the soft supersymmetry breaking Higgs and singlet 
masses are smaller than the Higgs doublet mass in the MSSM. 

In our analysis, we consider the limit with $M_{1}' \gg M_{1,2}$, where
$M_1'$, $M_1$, and $M_2$ are the gaugino masses for $U(1)'$, $U(1)_Y$, and 
$SU(2)_L$, respectively \cite{Erler:prd.66.015002, Han:hep-ph/0405244}. 
This limit suppresses the effects of the $U(1)'$ Higgs charges and greatly 
reduces the number of free parameters in the neutralino mass matrix.
The allowed mass range of the lightest 
neutralino then is limited ($m_{\chi^0} \alt 100$ GeV) compared to that of 
the MSSM, due to the large component of the singlino in the neutralino, 
which we assume is the LSP.
While the small mass of the LSP makes the heavy fermion and gauge boson channels of the MSSM irrelevant, 
it opens up new relevant channels, particularly the $Z$-resonance and the 
light Higgs resonance channels.
The lighter Higgs scalar ($h^0$) in this model is a mixture of the Higgs 
doublets and singlets, and it can be much lighter than the LEP bound of 
$m_{h^0} > 115$ GeV that applies to the SM-like Higgs bosons 
\cite{Erler:prd.66.015002, Han:hep-ph/0405244}.

In the MSSM, $Z$-resonance annihilation is not likely to be a relevant 
channel because the $Z \chi^0 \chi^0$ coupling is small when 
$\tan\beta \sim 1$. However, when the singlino component is introduced into 
the neutralino sector, the $Z \chi^0 \chi^0$ coupling can be enhanced to give 
sufficiently large $\chi^0 \chi^0$ annihilation.
This $Z$-resonance annihilation alone, with suitable parameter values, can 
reproduce the acceptable cold dark matter relic density in most of the allowed $\chi^0$ mass range in the $S$-model.

\section{Mass and Coupling of the Lightest Neutralino}

We consider a scenario with a massive $Z'$ and VEVs of 
$S_i$ ($i = 1, 2, 3$) that are large compared to other electroweak scale parameters.
There is an approximate decoupling of the neutralinos associated with 
the $Z'$ and the $S_i$, and the effective neutralino mass matrix for the 
remaining neutralinos in the basis of 
$\left\{\tilde{B},\tilde{W}_3,\tilde{H}_1^0,\tilde{H}_2^0,\tilde{S} \right\}$ 
is\footnote{In this limit, the neutralino matrix is basically the same as 
that of the one-singlet models and shares many of the same properties 
\cite{NMSSMneutralino, Menon:hep-ph/0404184}.}
\begin{eqnarray}
{\cal M}_{\chi^0}= 
\left( \matrix{M_1 & 0 & - g_1 v_1 / 2 & g_1 v_2 / 2 & 0 \cr
0 & M_2 & g_2 v_1 / 2 & - g_2 v_2 / 2 & 0 \cr
- g_1 v_1 / 2 & g_2 v_1 / 2 & 0 & - h_s s / \sqrt{2} & - h_s v_2 / \sqrt{2} \cr
g_1 v_2 / 2 & - g_2 v_2 / 2 & - h_s s / \sqrt{2} & 0 & - h_s v_1 / \sqrt{2} \cr
0 & 0 & - h_s v_2 / \sqrt{2} & - h_s v_1 / \sqrt{2} & 0 \cr} \right)
\label{eqn:massmatrix}
\end{eqnarray}
where 
$e = g_1\cos\theta_W = g_2\sin\theta_W$.
The VEVs of the Higgs doublets are $\langle H_i^0 \rangle \equiv 
\frac{v_i}{\sqrt{2}}$ with $\sqrt{v_1^2 + v_2^2} \simeq 246$ GeV and the VEV of the Higgs singlet is 
$\langle S^0 \rangle \equiv \frac{s}{\sqrt{2}}$. 
The mass eigenstates are ordered as 
$m_{\chi^0} = m_{\chi^0_1} < m_{\chi^0_2} < \cdots < m_{\chi^0_5}$. 
This mass matrix leads to a kind of see-saw mechanism 
\cite{Hesselbach:epjc.23.149, Erler:prd.66.015002, Menon:hep-ph/0404184} 
that makes the lightest neutralino mass very small.
In our analysis we scanned over the parameters
\begin{eqnarray}
\tan\beta &=& 1.03, 1.5, 2.0, 2.5 \nonumber \\
h_s &=& 0.1~\mbox{to 0.75 \ (in steps of 0.01)} \nonumber \\
M_2 &=& -500~\mbox{GeV to 500~GeV \ (in steps of 1 GeV)} \nonumber \\
s &=& 100~\mbox{GeV to 500~GeV \ (in steps of 1 GeV)} \nonumber
\end{eqnarray}
We use the gaugino mass unification relations ($M_1 = \frac{5}{3} \frac{g_1^2}{g_2^2} M_2 \simeq 0.5 M_2$), but take $M_1' \gg M_2$. 
The $U(1)'$ charge dependence vanishes from the effective mass matrix for 
large $M_1'$.

The MSSM mass matrix\footnote{The MSSM parameter range considered here (especially for $\tan \beta$) is not experimentally allowed (e.g. by the lightest Higgs mass), 
but is chosen to demonstrate the effects of adding the $U(1)'$.} 
corresponds to dropping the last row and column in the matrix of 
Eq. (\ref{eqn:massmatrix}) and taking $\mu = h_s s / \sqrt{2}$.
The MSSM upper bound on $m_{\chi^0}$ is very sensitive to the value of $M_2$, 
while that of the $S$-model depends sensitively on the value of $h_s$.
In both models, the upper bound on the lightest neutralino mass has its largest value when $\tan\beta$ is $1.03$ (or $1$)\footnote{Since the exact 
value $\tan\beta = 1$ would make the coupling of $Z \chi^0 \chi^0$ vanish 
in both the MSSM and the $S$-model, we choose $\tan\beta = 1.03$ instead 
of $1$.}.
When $\tan\beta \sim 1$, the $m_{\chi^0}$ bound in the $S$-model increases almost linearly with $h_s$. 
We choose our upper bound of $h_s$ to be $0.75$ as advocated in Ref.~\cite{Erler:prd.66.015002} to keep the running of $h_s$ finite up to 
Plank scale \cite{Menon:hep-ph/0404184, Erler:prd.66.015002}.

We find that
\begin{equation}
m_{\chi^0} \alt 99 \; \mbox{GeV}.
\label{eqn:upperbound}
\end{equation}
The maximum value of $m_{\chi^0}$ occurs with $\tan\beta = 1.03$, $h_s = 0.75$, $M_2 = -183$ GeV and $s = 130$ GeV for the $S$-model\footnote{The upper bound on $m_{\chi^0}$ in the different limits are: 
(i) for $M_1' \gg M_2$, $s_i \sim {\cal O}$(EW), $m_{\chi^0} \alt 100$ GeV; 
(ii) for $M_1' = M_1$, $s_i \gg {\cal O}$(EW), $m_{\chi^0} \alt 280$ GeV; 
(iii) for $M_1' = M_1$, $s_i \sim {\cal O}$(EW), $m_{\chi^0} \alt 100$ GeV. 
We scanned $100$ to $1000$ GeV for $M_2$ and $s_i$ and assumed $h_s \alt 0.75$ with 
$\lambda_s \alt 0.2$. 
For the $U(1)'$ couplings, various $E_6$ model charge assignments 
$(\chi, \psi, \eta)$ and the GUT motivated coupling constant 
$g_{Z'} = \sqrt{\frac{5}{3}} g_1$ were assumed. 
For a full $9 \times 9$ neutralino mass matrix, see 
Ref. \cite{Erler:prd.66.015002, Han:hep-ph/0405244}.}
while for the MSSM the maximum $m_{\chi^0} \alt 254$ GeV occurs for $\tan\beta = 1.03$, $M_2 = -500$ GeV and $\mu \simeq 261$ GeV.

The small LSP mass makes the $Z$-resonance contribution to the relic density relevant in the $S$-model. For $Z\chi_i^0\chi_j^0$ and $Z'\chi_i^0\chi_j^0$ interactions, 
the general Lagrangian is given by \cite{Hesselbach:epjc.23.149}
\begin{eqnarray}
{\cal L}&=& \frac{1}{4} g_Z Z_\mu \left( \bar{\tilde{H_1^0}} \gamma^\mu \gamma^5 \tilde{H_1^0} - \bar{\tilde{H_2^0}} \gamma^\mu \gamma^5 \tilde{H_2^0} \right) \nonumber \\
&+& \frac{1}{4} g_{Z'} Z'_\mu \left( 2Q_1 \bar{\tilde{H_1^0}} \gamma^\mu \gamma^5 \tilde{H_1^0} + 2Q_2 \bar{\tilde{H_2^0}} \gamma^\mu \gamma^5 \tilde{H_2^0} + 2Q_S \bar{\tilde{S}} \gamma^\mu \gamma^5 \tilde{S} \right)
\end{eqnarray}
which, in terms of neutralino mass eigenstates, can be written as
\begin{eqnarray}
{\cal L}_{Z(Z') \chi^0 \chi^0}&=& \frac{1}{2} g_Z Z_\mu \bar{\chi}^0_i \gamma^\mu \left( O''^L_{ij} P_L + O''^R_{ij} P_R \right) \chi^0_i \nonumber \\
&+& \frac{1}{2} g_{Z'} Z'_\mu \bar{\chi}^0_i \gamma^\mu \left( I''^L_{ij} P_L + I''^R_{ij} P_R \right) \chi^0_i \,.
\end{eqnarray}
Here the couplings are defined as
\begin{eqnarray}
O''^L_{ij} 
& \equiv & -\frac{1}{2} N_{i3} N_{j3}^* + \frac{1}{2} N_{i4} N_{j4}^*
\;\;\;\;\;
O''^R_{ij} \equiv  \frac{1}{2} N_{i3}^* N_{j3} - \frac{1}{2} N_{i4}^* N_{j4} \\
I''^L_{ij} 
& \equiv & -Q_1 N_{i3} N_{j3}^* - Q_2 N_{i4} N_{j4}^* - Q_S N_{i5} N_{j5}^* \\
I''^R_{ij} 
& \equiv & Q_1 N_{i3}^* N_{j3} + Q_2 N_{i4}^* N_{j4} + Q_S N_{i5}^* N_{j5},
\end{eqnarray}
where the matrix $N$ relates the weak and mass bases.
$Q_1$, $Q_2$ and $Q_S$ are the $U(1)'$ charge of the $H_1^0$, $H_2^0$ and $S$, respectively.
The $Z\chi^0 \chi^0$ coupling 
$O''^L_{11} = - O''^R_{11}
= -\frac{1}{2} \left( |N_{13}|^2 - |N_{14}|^2 \right)$ is composed of the two doublet Higgsino components coupling to the $Z$ boson.
A massive (TeV-scale) $Z'$ would not provide sufficiently large annihilation for our light neutralinos, and thus in our analysis we use only the $Z$-resonance annihilation.


\begin{figure}[htb]
\begin{minipage}[c]{\textwidth}
\begin{minipage}[b]{.49\textwidth}
\centering\leavevmode
\epsfxsize=3in
\epsfbox{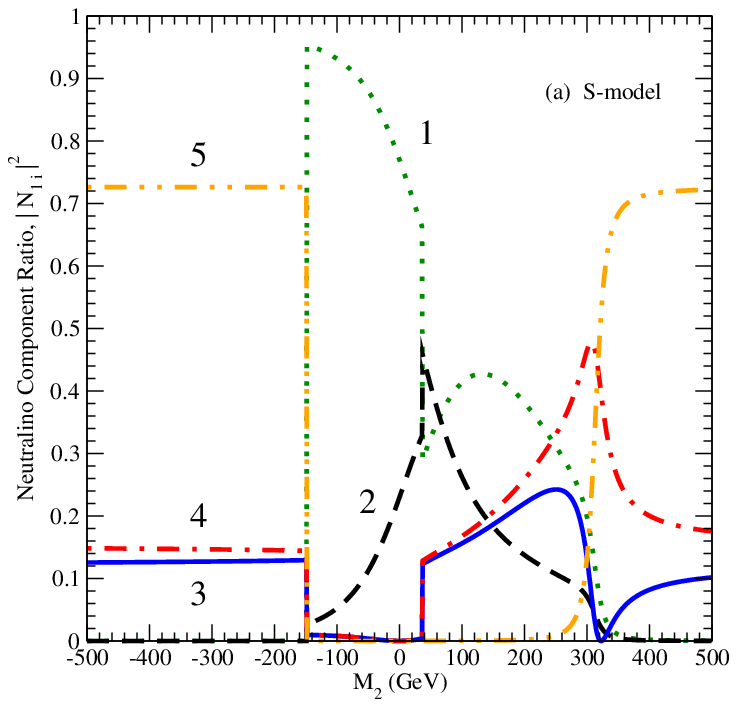}
\end{minipage}
\hfill
%
\begin{minipage}[b]{.49\textwidth}
\centering\leavevmode
\epsfxsize=3in
\epsfbox{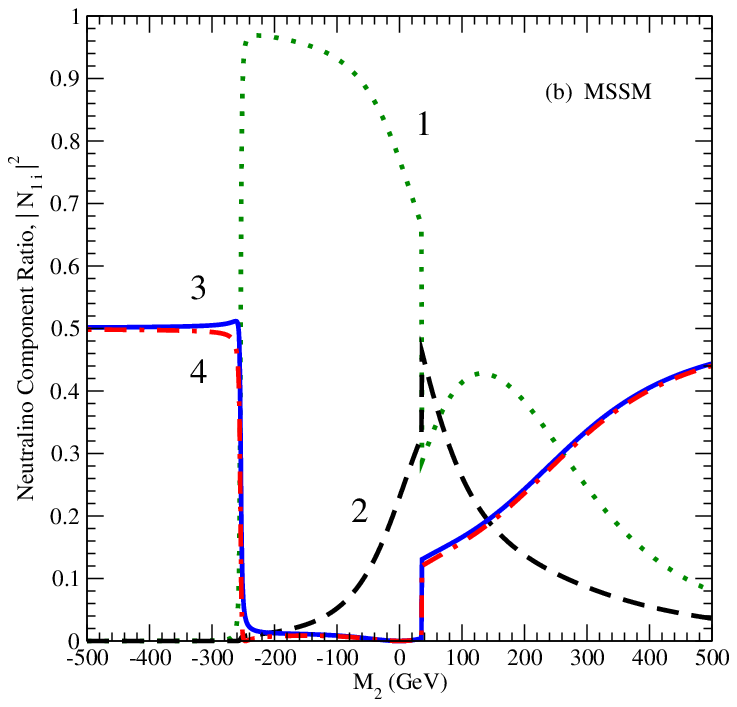}
\end{minipage} 
\end{minipage}
\caption{
The fraction of the lightest neutralino
($|N_{1i}|^2$) versus $M_2$ for (a) the $S$-model and (b) the MSSM.  
All components are presented: 
$|N_{11}|^2$ ($\tilde B$, dot), $|N_{12}|^2$ ($\tilde W_3$, dash), $|N_{13}|^2$ ($\tilde H_1^0$, solid), $|N_{14}|^2$ ($\tilde H_2^0$, dash-dot), and $|N_{15}|^2$ ($\tilde S$, dash-dot-dot).
The curve for $|N_{1i}|^2$ is labeled by $i$.
Fixed values of $h_s = 0.75$, $\tan\beta = 1.03$ and $s = 250$ GeV are illustrated.
The difference of $|N_{13}|^2$ and $|N_{14}|^2$ is large when the singlino component is present even though $\tan\beta \sim 1$.}
\label{fig:component}
\end{figure}

Figure \ref{fig:component} presents the $|N_{1i}|^2$ with $i = $ 1 
($\tilde B$, dot), 2 ($\tilde W_3$, dash), 3 ($\tilde H_1^0$, solid), 
4 ($\tilde H_2^0$, dash-dot), 5 ($\tilde S$, dash-dot-dot), 
for $M_2 = -500$ to 500 GeV and a fixed set of values $h_s = 0.75$, 
$\tan\beta = 1.03$ and $s = 250$ GeV.
The value of each $|N_{1i}|^2$ is shown versus $M_2$ in 
(a) the $S$-model and (b) the MSSM.
We note that in the region around $M_2 \sim 300$ GeV the singlino component 
$|N_{15}|^2$ increases, and $|N_{13}|^2$ deviates substantially 
from $|N_{14}|^2$ making the $Z\chi^0 \chi^0$ coupling large.
This does not happen in the MSSM for the same parameters.
In addition, for $M_2 \alt -150$ GeV, the singlino dominates and the 
difference of $|N_{13}|^2$ and $|N_{14}|^2$ in the $S$-model is larger 
than that in the MSSM.
As an example, the components of the LSP for $M_2 = 330$ GeV are
\begin{eqnarray}
\chi^0 &=& 0.18 {\tilde B} - 0.11 {\tilde W_3} - 0.09 {\tilde H_1^0} -0.58 {\tilde H_2^0} + 0.78 {\tilde S} \mbox{~~($S$-model),} \\
\chi^0 &=& 0.46 {\tilde B} - 0.27 {\tilde W_3} + 0.60 {\tilde H_2^0} - 0.60 {\tilde H_3^0} \mbox{~~~~~~~~~~~~~(MSSM).}
\end{eqnarray}

The large difference in $|N_{13}|^2$ and $|N_{14}|^2$ for the $S$-model is remarkable in view of the fact that $\tan\beta = 1$ makes the $\tilde H^0_1$ and $\tilde H^0_2$ parts 
in the mass matrix of Eq.~(\ref{eqn:massmatrix}) the same up to the sign, 
which leads one to expect $|N_{13}|^2 - |N_{14}|^2 \sim 0$, as is the case in 
the MSSM (Figure~\ref{fig:component} (b)).
Thus, the addition of the singlino component plays a critical role in 
enhancing the difference of $|N_{13}|^2$ and $|N_{14}|^2$, which is 
particularly important for the $\chi^0$ in the $S$-model to generate 
sufficient annihilation mediated by the $Z$ boson.
The suppressed coupling (by $\tan\beta \sim 1$) can still be large enough to 
give an acceptable relic density.
Figure \ref{fig:component} also shows that the gaugino components 
$|N_{11}|^2$ (bino) and $|N_{12}|^2$ (wino), especially the bino, are 
dominant for relatively small values of $|M_2|$ in both models.


\begin{figure}[htb]
\begin{minipage}[c]{\textwidth}
\begin{minipage}[b]{.32\textwidth}
\centering\leavevmode
\epsfxsize=2.2in
\epsfbox{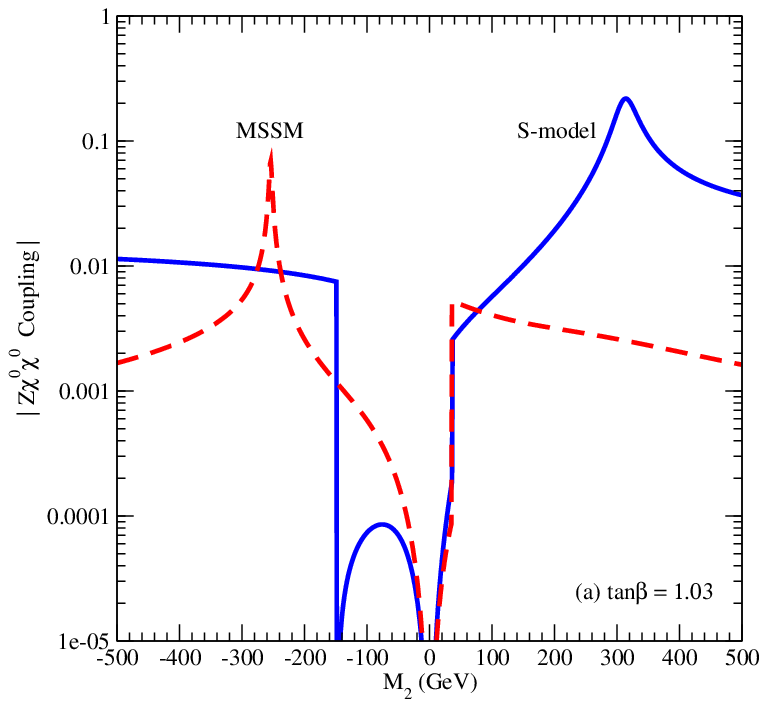}
\end{minipage}
\hfill
\begin{minipage}[b]{.32\textwidth}
\centering\leavevmode
\epsfxsize=2.2in
\epsfbox{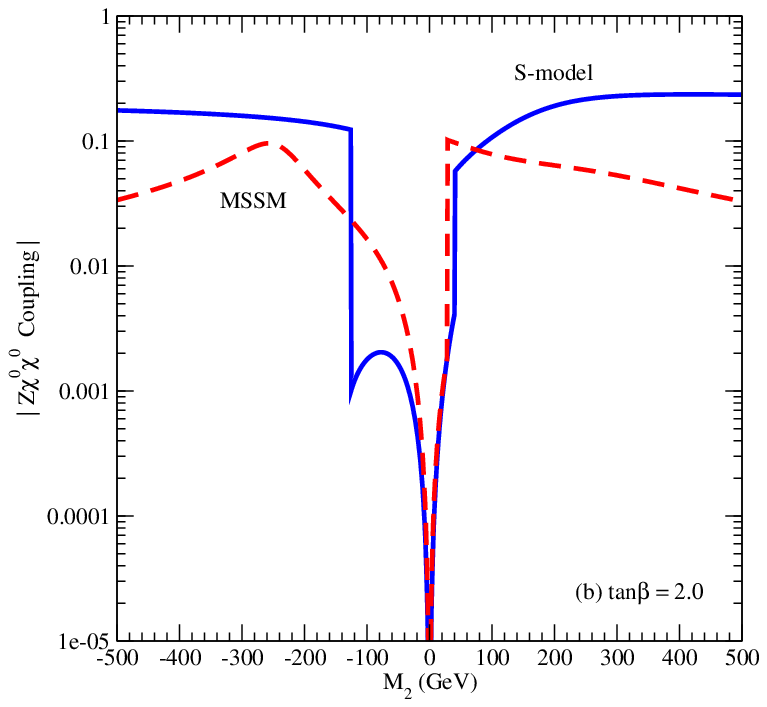}
\end{minipage}
\hfill
\begin{minipage}[b]{.32\textwidth}
\centering\leavevmode
\epsfxsize=2.2in
\epsfbox{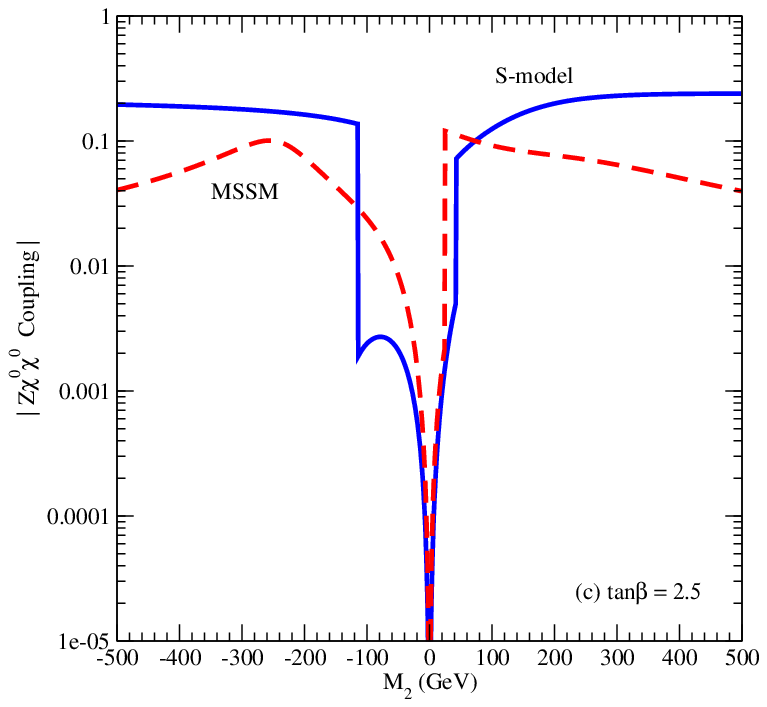}
\end{minipage}
\end{minipage}
\caption{
The $Z\chi^0\chi^0$ coupling $|O''_{11}|$ versus $M_2$ for 
(a) $\tan\beta = 1.03$, (b) $\tan\beta = 2.0$, and (c) $\tan\beta = 2.5$, 
in the $S$-model (solid) and in the MSSM (dash).
Fixed values of $h_s = 0.75$ and $s = 250$ GeV are used.
The coupling is much larger in the $S$-model than in the MSSM in most of the parameter space.}
\label{fig:coupling}
\end{figure}

We present $|O''_{11}| \equiv |O''^L_{11}| = |O''^R_{11}|$ for the $S$-model and the MSSM in Figure \ref{fig:coupling}.
For the same values of $h_s = 0.75$ and $s = 250$ GeV, we select 3 different 
$\tan\beta$ values of (a) $\tan\beta = 1.03$, (b) $\tan\beta = 2.0$, and (c) $\tan\beta = 2.5$.
For each $\tan\beta$, the $Z \chi^0 \chi^0$ coupling is 
much larger in the $S$-model than in the MSSM in most of the parameter 
space. We numerically checked that this feature holds for other choices 
of the parameter values ($0.4 \alt h_s  \alt 1.0$, $100$ GeV 
$\alt s \alt 1000$ GeV, $0.5 \alt \tan\beta \alt 2.5$).

For $\tan\beta \sim 1$ (Figure \ref{fig:coupling} (a)), the $Z\chi^0 \chi^0$ 
coupling in the $S$-model is at its maximum for $M_2 \simeq 300$ GeV and there is a relatively small but still noticeable peak in the MSSM at $M_2 \simeq -250$ GeV.
The $\chi^0$ relic density depends on not only the coupling but also whether $m_{\chi^0}$ is near $M_Z/2$.


\begin{figure}[htb]
\begin{minipage}[c]{\textwidth}
\begin{minipage}[b]{.48\textwidth}
\centering\leavevmode
\epsfxsize=3in
\epsfbox{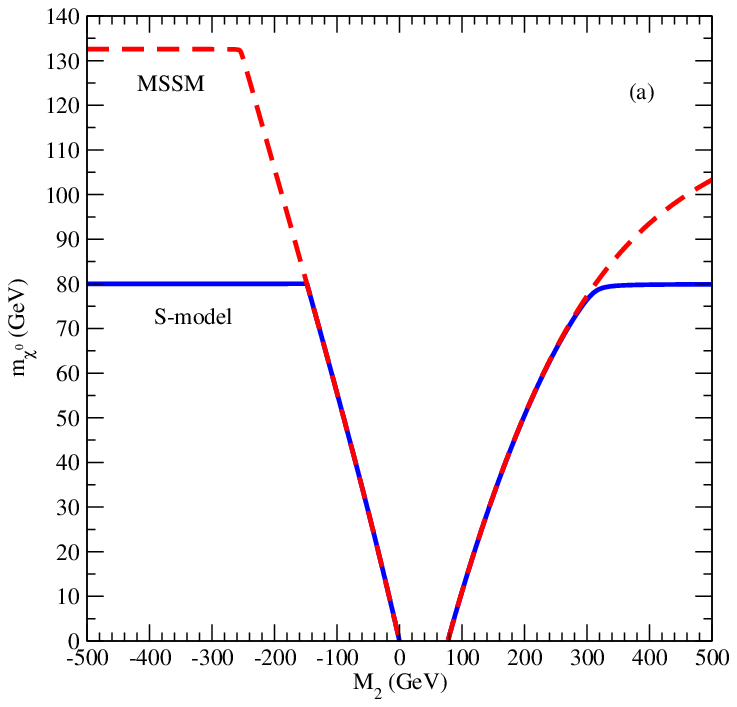}
\end{minipage}
\hfill
%
\begin{minipage}[b]{.48\textwidth}
\centering\leavevmode
\epsfxsize=3in
\epsfbox{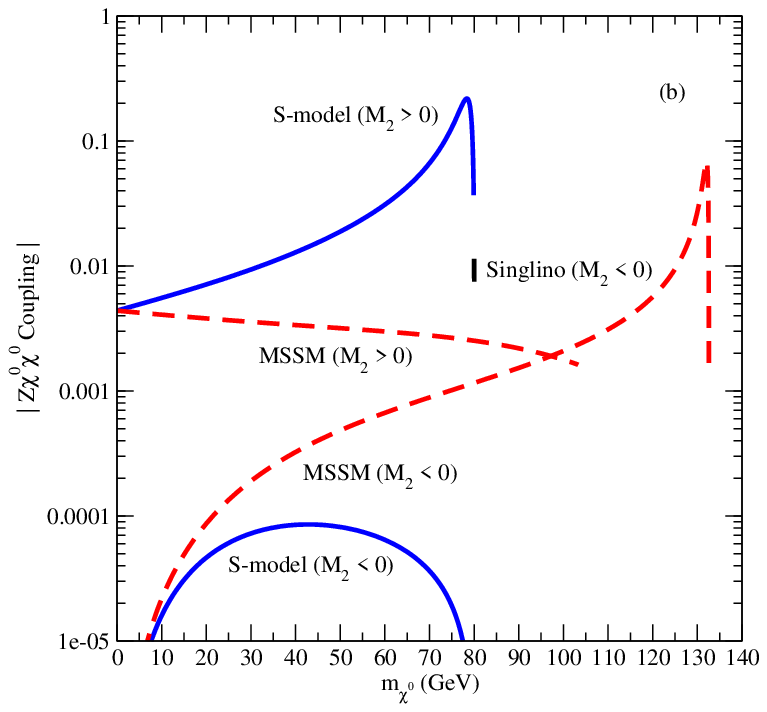}
\end{minipage} 
\end{minipage}
\caption{
(a) The lightest neutralino mass as a function of $M_2$ and 
(b) the $Z\chi^0 \chi^0$ coupling ($|O''_{11}|$) versus the lightest 
neutralino mass in the $S$-model (solid) and the MSSM (dash).
Fixed values of $h_s = 0.75$, $\tan\beta = 1.03$ and $s = 250$ GeV are used.
The $S$-model has a smaller $m_{\chi^0}$ bound, and, for $M_2 > 0$, larger 
$Z \chi^0 \chi^0$ coupling than the MSSM.}
\label{fig:mass}
\end{figure}

Figure \ref{fig:mass} shows (a) the $\chi^0$ mass versus $M_2$ and 
(b) $|O''_{11}|$ versus the LSP mass.
The same parameter values as Figures \ref{fig:component} 
and \ref{fig:coupling} (a) are used.
The LSP mass in the $S$-model is almost constant for $M_2 \alt -200$ GeV 
and $M_2 \agt 300$ GeV, and it is smaller than in the MSSM.
The lightest neutralino masses in both models have practically identical 
dependence on $M_2$ before they reach the flat curves in our parameter 
choice $(\tan\beta \simeq 1)$.
Enhancements of couplings in the $S$-model and the MSSM  
are found around $m_{\chi^0} \sim 80$ GeV and $130$ GeV, respectively, 
that is, for the flat parts of the mass curves in Figure \ref{fig:mass} (a).

\section{Neutralino Annihilation Mediated by the $Z$ Boson}

The relic density of the lightest neutralino
is inversely proportional to the annihilation cross section
of $\chi^0$ pairs. When kinematically allowed, the neutralino pairs 
annihilate into pairs of fermions, gauge bosons, and Higgs bosons 
through $s$, $t$, and $u$ channel diagrams. In general, the annihilation 
cross section is greatly enhanced by the $Z$ boson or the Higgs boson 
($\phi^0$) poles, and the relic density is correspondingly suppressed when 
$2 m_{\chi^0} \sim M_Z$ or $2 m_{\chi^0} \sim M_\phi$. 

The Higgs masses and couplings in the $S$-model depend on 
several free parameters.
Moreover, the light Higgs width is very narrow and a fine-tuning of parameters would be required for neutralino annihilation through $\phi^0$ 
to generate the appropriate relic density, and the other Higgs bosons 
would often be too heavy to have resonance effects because the LSP is light \cite{Han:hep-ph/0405244}.
We are therefore interested in the case that the 
$Z\chi^0\chi^0$ coupling is sizable and $Z$ mediated annihilation alone 
can lead to an acceptable relic density ($\Omega_{\rm CDM} h^2$).
A more complete investigation is in progress to include all Higgs resonance effects.
We require a lower bound on $m_{\chi^0}$ of $M_Z / 2$ so that 
$Z \to \chi^0\chi^0$ does not affect the $Z$ width significantly\footnote{
Since $\Delta \Gamma_{\rm inv} < 2.0$ MeV (95\% C.L.) \cite{invwidth}, 
smaller $m_{\chi^0}$ is possible \cite{Menon:hep-ph/0404184}. 
For reasonably smaller masses, our results are unchanged.}.
Then the lightest neutralino mass in the $S$-model is constrained to the range
\begin{eqnarray}
46 \mbox{ GeV} \alt m_{\chi^0} \alt 100 \mbox{ GeV}. \label{eqn:masslimit}
\end{eqnarray}
with the upper bound from Eq. (\ref{eqn:upperbound}).

We evaluate the annihilation cross section including interferences among 
all diagrams by calculating the amplitude of each diagram with the helicity 
amplitude formalism, then numerically evaluate the full matrix element squared.
The neutralino relic density is calculated with relativistic Boltzmann 
averaging, neutralino annihilation threshold effects
and Breit-Wigner poles \cite{Baer&Brhlik, Barger&Kao, Roszkowski}.


\begin{figure}[htb]
\begin{minipage}[c]{\textwidth}
\begin{minipage}[b]{.32\textwidth}
\centering\leavevmode
\epsfxsize=2.2in
\epsfbox{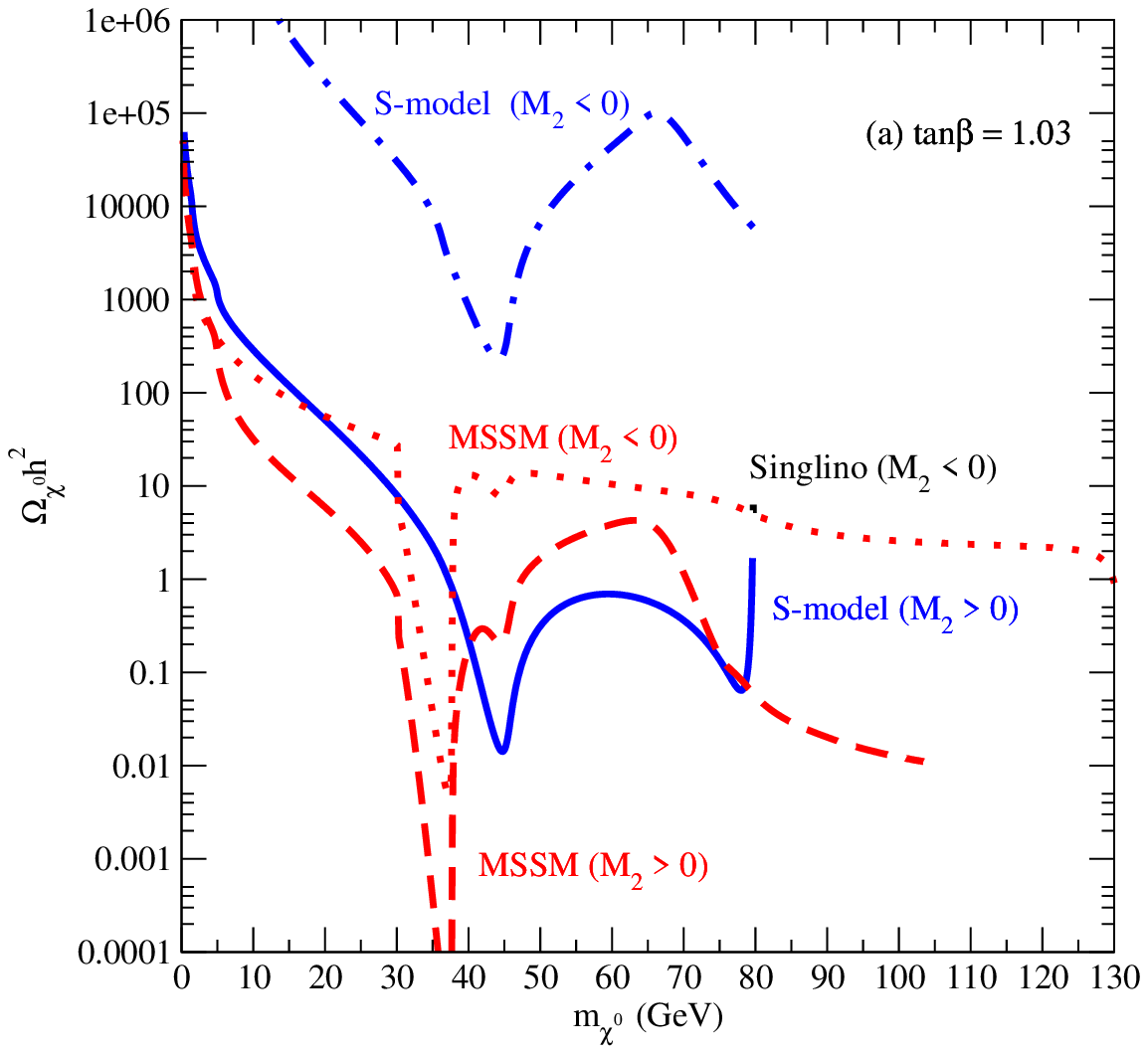}
\end{minipage}
\hfill
\begin{minipage}[b]{.32\textwidth}
\centering\leavevmode
\epsfxsize=2.2in
\epsfbox{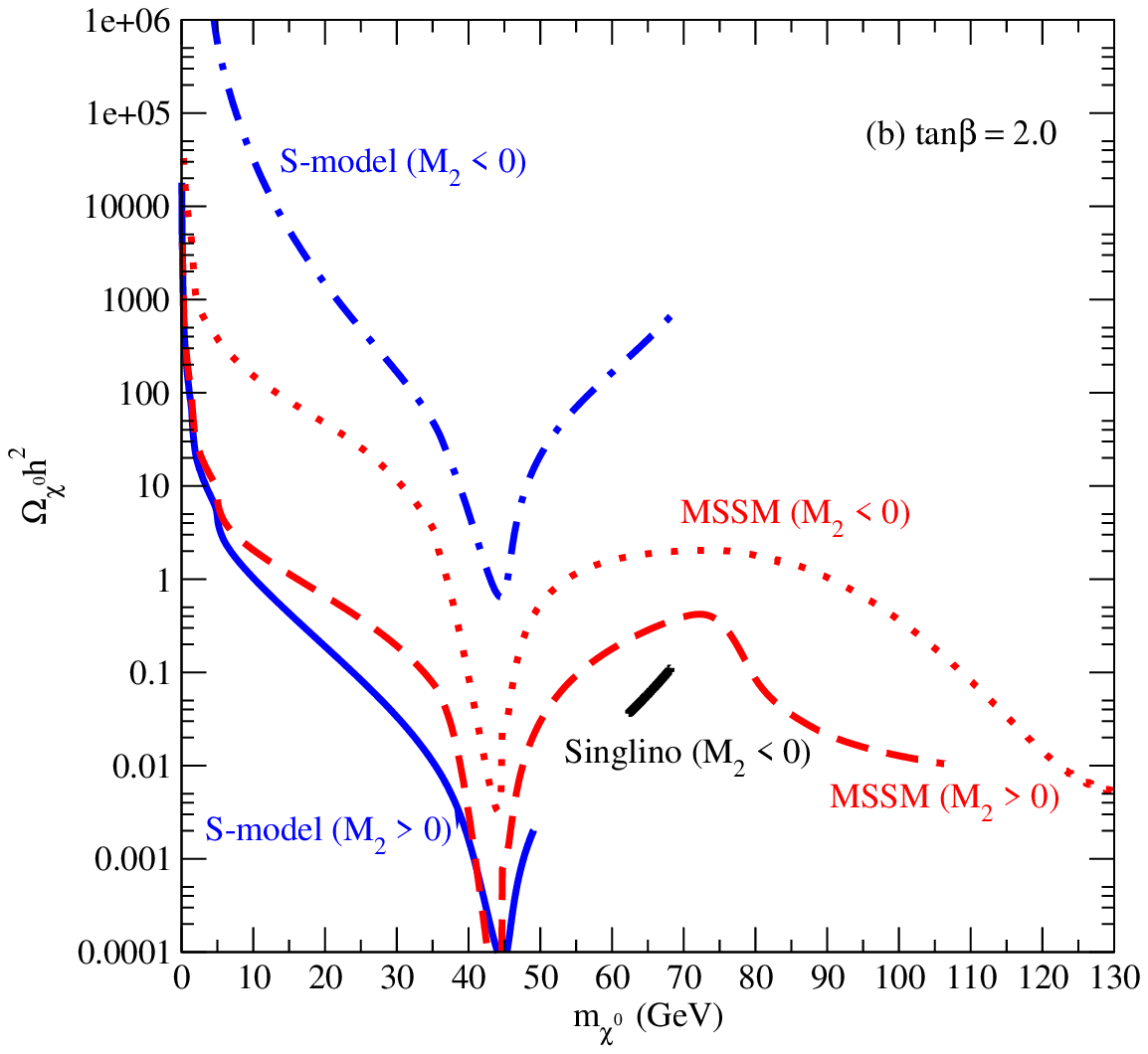}
\end{minipage}
\hfill
\begin{minipage}[b]{.32\textwidth}
\centering\leavevmode
\epsfxsize=2.2in
\epsfbox{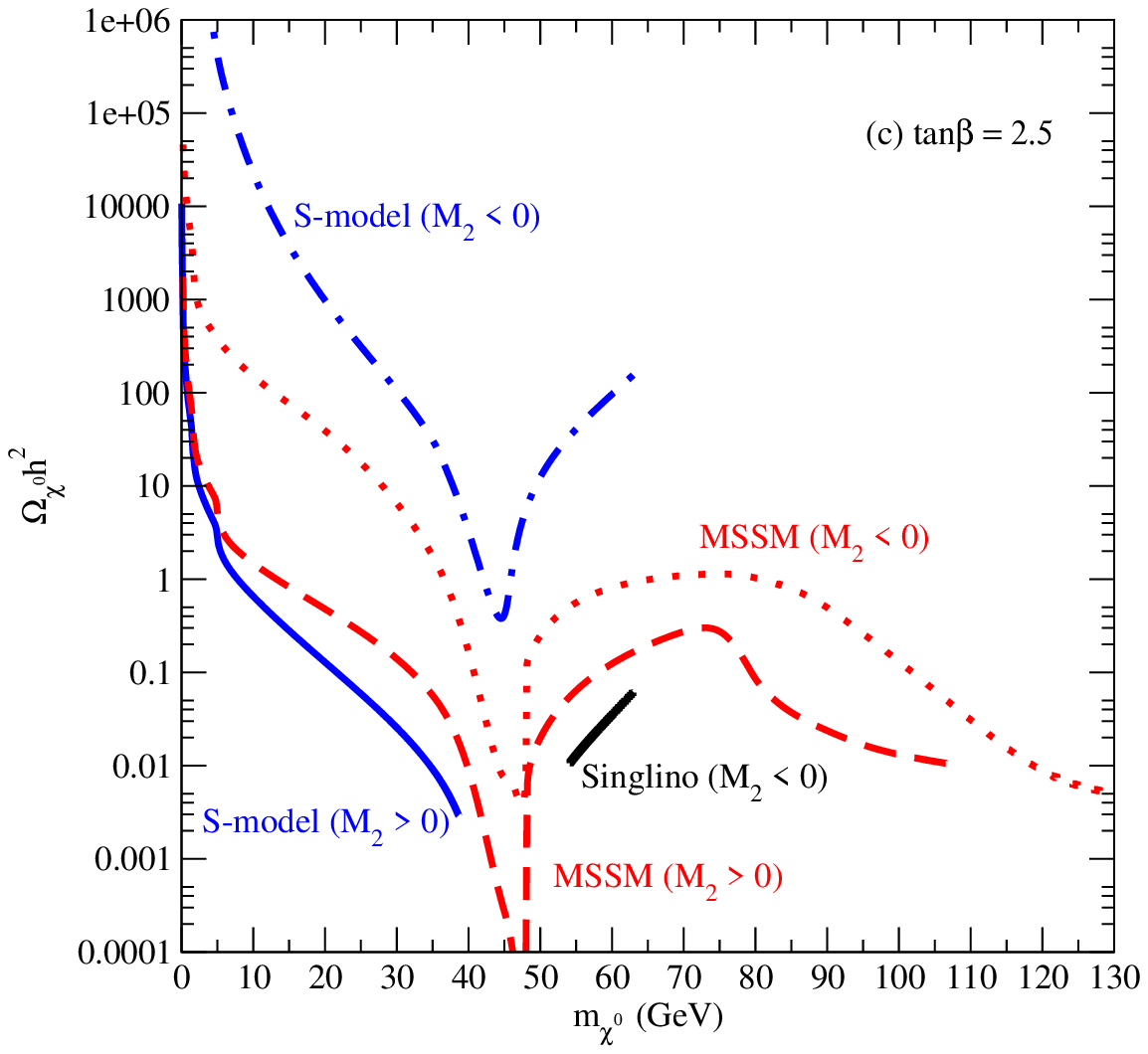}
\end{minipage}
\end{minipage}
\caption{
The neutralino relic density ($\Omega_{\chi^0} h^2$) versus $m_{\chi^0}$
in the $S$-model [$M_2 > 0$ (solid) and $M_2 < 0$ (dash-dot)]
and the MSSM [$M_2 > 0$ (dash) and $M_2 < 0$ (dot)] for 
(a) $\tan\beta = 1.03$, (b) $\tan\beta = 2.0$, and (c) $\tan\beta = 2.5$. 
The isolated black curve corresponds to the singlino-dominated flat part of the $S$-model in Figure \ref{fig:mass} with $M_2 < 0$.
Fixed values of $h_s = 0.75$ and $s = 250$ GeV are used.
The acceptable relic density is $\Omega_{\chi^0} h^2 \sim 0.1$~.}
\label{fig:relic}
\end{figure}

We show the neutralino relic density ($\Omega_{\chi^0} h^2$) versus $m_{\chi^0}$ in Figure \ref{fig:relic} for both the $S$-model\footnote{
For similar relic density results in the nMSSM, see Ref. 
\cite{Menon:hep-ph/0404184}. For the NMSSM relic density calculations, see Ref. \cite{NMSSMrelic}.} and the MSSM for the $\tan\beta$ choices 
(a) $\tan\beta = 1.03$, (b) $\tan\beta = 2.0$, and (c) $\tan\beta = 2.5$.
We included all possible channels for the MSSM calculation; for the $S$-model we included only the kinematically relevant $s$-channel processes with the $Z \chi^0 \chi^0$ coupling, $\chi^0 \chi^0 \to Z \to f_i \bar f_i$, $W^+ W^-$ ($f_i$ is a SM fermion).
We scanned over $-500$ GeV $< M_2 <$ 500 GeV for $h_s = 0.75$, and $s = 250$ GeV to calculate $m_{\chi^0}$ and evaluate the $Z \chi^0 \chi^0$ coupling.
There are several interesting aspects to note.
\begin{itemize}
\item When $m_{\chi^0} \simeq M_Z/2$, there is a dip in the 
curve of relic density versus $m_{\chi^0}$ since there is a peak in 
the annihilation cross section enhanced by the $Z$-pole.
\item In the $S$-model, a positive value of $M_2$ gives two acceptable CDM density regions (around the $Z$ pole and the enhanced coupling region such as $m_{\chi^0} \sim 80$ GeV in the $\tan\beta = 1.03$ case) while a negative $M_2$ leads to $\Omega_{\chi^0} h^2 \agt 1$ except for the singlino-dominated region.
\item The MSSM can also give an acceptable CDM density, but the model is excluded for $\tan\beta \sim 1$ by the LEP Higgs bound, owing to its small Higgs mass.
The $S$-model Higgs contains a singlet component and the light Higgs can be compatible with the LEP constraint \cite{Han:hep-ph/0405244}.
\item The small isolated region for relatively large $m_{\chi^0}$ corresponds to $M_2 < 0$ 
with the lightest neutralino being singlino-dominated in the $S$-model\footnote{The case in which the lightest neutralino is mostly 
singlino-like (e.g., $\tan\beta = 1.03$, $m_{\chi^0} \sim 80$ GeV for both 
$M_2 < 0$ and $M_2 > 0$), is qualitatively similar to those studied in 
Ref. \cite{deCarlos:plb.407.12}.}.
As observed in Figure \ref{fig:component}, a sudden sizable deviation of $|N_{13}|^2$ from $|N_{14}|^2$ in a singlino dominated region ($M_2 \alt -150$ GeV for $\tan\beta = 1.03$) provides a sudden drop of $\Omega_{\chi^0} h^2$ from the rest of the $M_2 < 0$ curve.
\end{itemize}


\begin{figure}[htb]
\begin{minipage}[c]{\textwidth}
\begin{minipage}[b]{.32\textwidth}
\centering\leavevmode
\epsfxsize=2.2in
\epsfbox{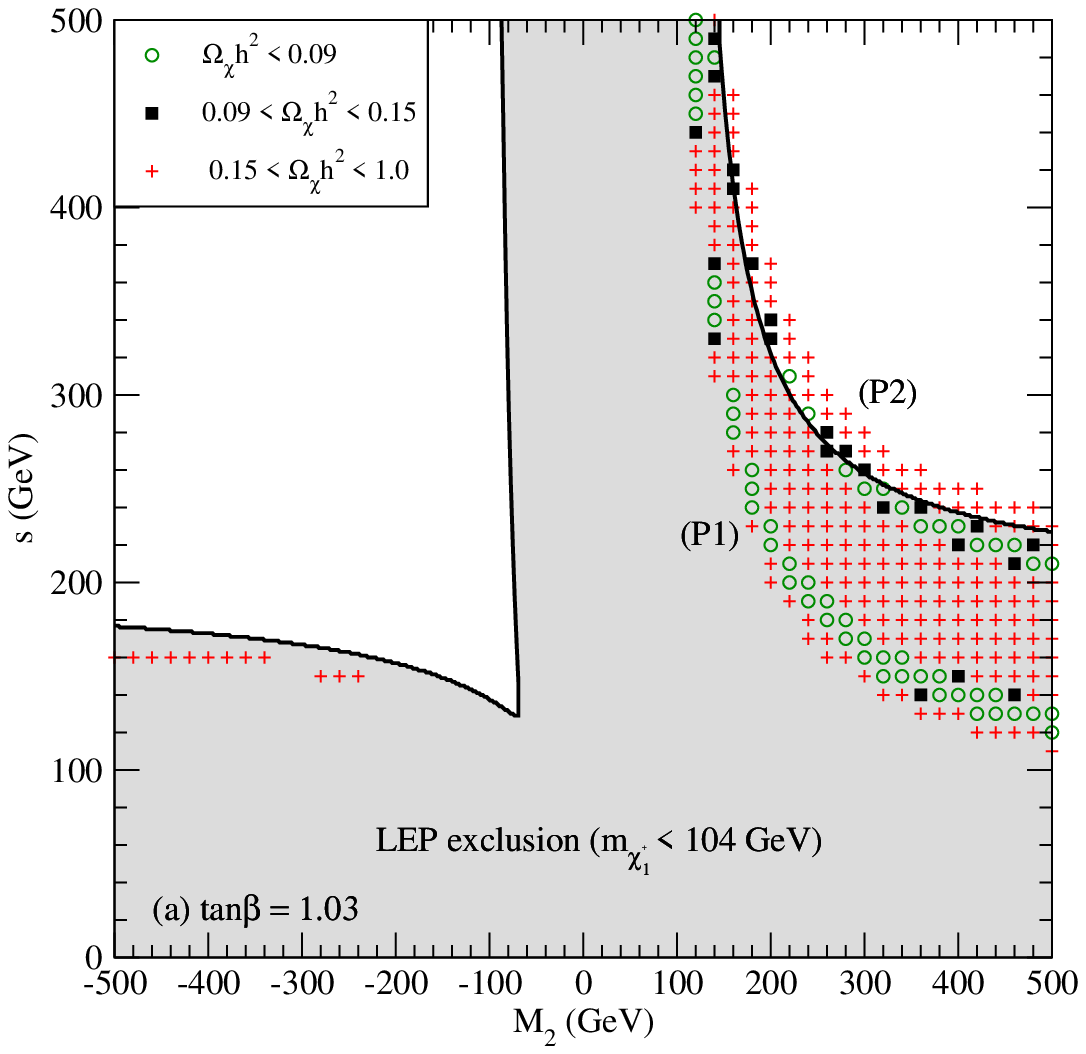}
\end{minipage}
\hfill
\begin{minipage}[b]{.32\textwidth}
\centering\leavevmode
\epsfxsize=2.2in
\epsfbox{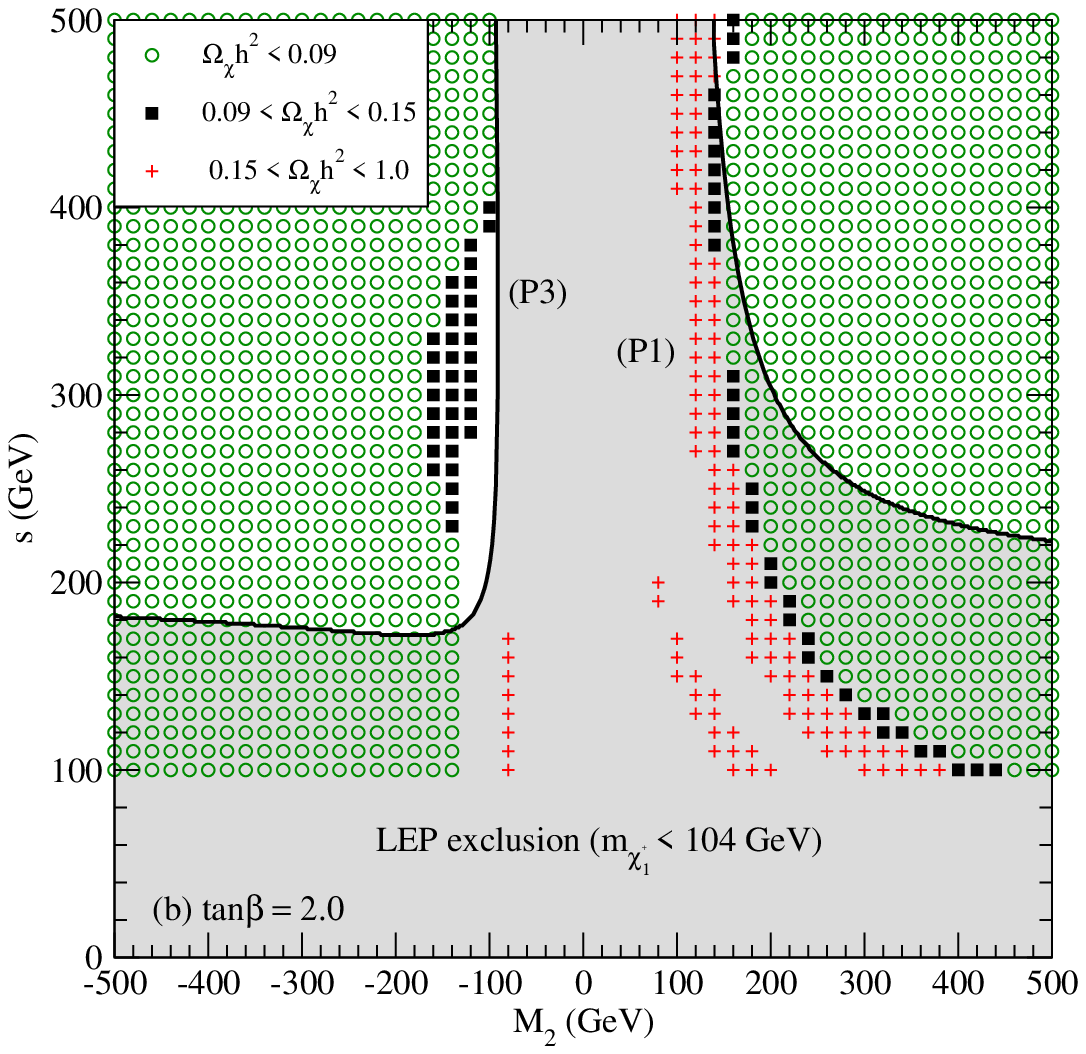}
\end{minipage}
\hfill
\begin{minipage}[b]{.32\textwidth}
\centering\leavevmode
\epsfxsize=2.2in
\epsfbox{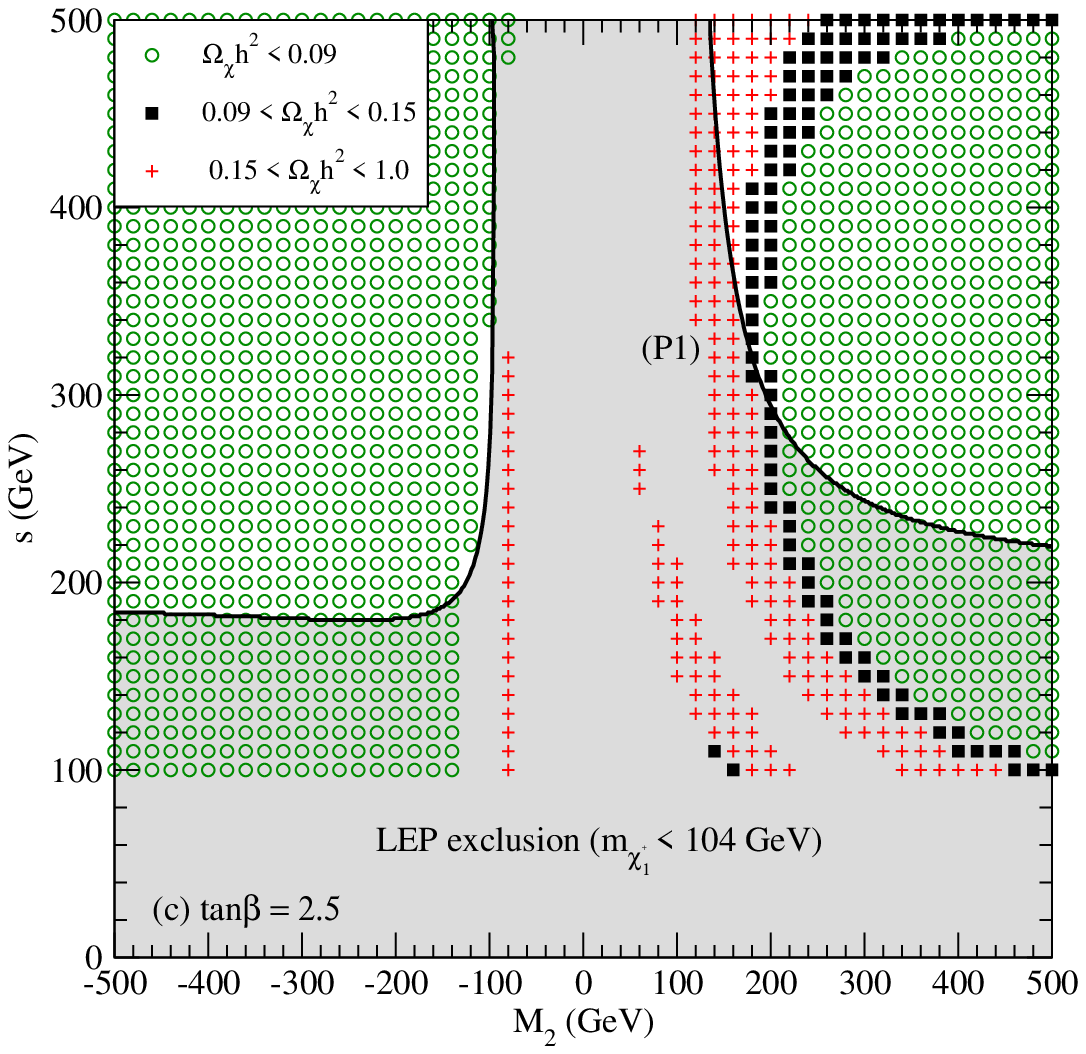}
\end{minipage}
\end{minipage}
\caption{
Regions of the $S$-model neutralino relic density in the $M_2$-$s$ plane (with $s$ scanned only above $100$ GeV) for $0.09 < \Omega_{\chi^0} h^2 < 0.15$ (filled square; $3 \sigma$ allowed range), $\Omega_{\chi^0} h^2 < 0.09$ (open circle), and $0.15 < \Omega_{\chi^0} h^2 < 1.0$ (cross).
Three representative values of $\tan\beta$ are chosen: 
(a) $\tan\beta = 1.03$, (b) $\tan\beta = 2.0$, and (c) $\tan\beta = 2.5$, and a fixed value of $h_s = 0.75$ is used. 
The shaded region of the parameter space (bounded by solid curves) is 
excluded by the LEP 2 chargino mass limit ($m_{\chi^+_1} \alt 104$ GeV).
There exist sizable regions (filled square) in the parameter 
space consistent with the relic density constraint outside of the chargino mass exclusion boundary.}
\label{fig:contour}
\end{figure}

Figure \ref{fig:contour} presents ranges of neutralino relic density 
in regions of the $M_2$-$s$ plane in the $S$-model. 
We choose $h_s = 0.75$ with (a) $\tan\beta = 1.03$, (b) $\tan\beta = 2.0$, 
and (c) $\tan\beta = 2.5$.
Also shown is the  region excluded by the LEP 2 chargino mass bound 
with $m_{\chi^+_1} < 104$ GeV \cite{Abbiendi:epjc.14.187}
The parameter points at which the light chargino is the LSP were omitted.
The tree level mass for the lighter chargino ($m_{\chi^+_1} < m_{\chi^+_2}$) 
is evaluated with the chargino mass matrix
\begin{equation}
m_{\chi^\pm} =
\left(
\begin{array}{cc}
M_2 & \sqrt{2} M_W \sin\beta \\
\sqrt{2} M_W \cos\beta & \mu_{\rm eff}
\end{array}
\right).
\end{equation}

We present the $S$-model neutralino relic density in 3 regions: 
$0.09 < \Omega_{\chi^0} h^2 < 0.15$ (filled square), 
$\Omega_{\chi^0} h^2 < 0.09$ (open circle), and 
$0.15 < \Omega_{\chi^0} h^2 < 1.0$ (cross).
The $3 \sigma$ range\footnote{Since we are using tree level masses and couplings, we allow rather conservative $3 \sigma$ range for the allowed CDM relic density.} of the CDM relic density of Eq. (\ref{eqn:sdsswmap}) is $0.09 < \Omega_{\chi^0} h^2 < 0.15$; however if there are other sources of dark matter in addition to the lightest neutralino, the range $\Omega_{\chi^0} h^2 < 0.09$ would be relevant.
Due to the finite grid, the filled square should be understood to be on the boundary of the open circle and the cross.

There are three separate regions that have an acceptable CDM density (filled square): 
(P1) Near the $Z$ pole, (P2) enhanced coupling region ($\tan\beta \approx 1$ case), and (P3) isolated singlino region.
In general, the $Z \chi^0 \chi^0$ coupling is enhanced by the singlino component.
There appears a sudden peak of the enhancement (P2) when $\tan\beta \approx 1$ as we can see from Figure \ref{fig:coupling} (a).
In the enhanced coupling region (P2), even when the $\chi^0$ mass is significantly distant from the $Z$ pole, an acceptable relic density can be obtained.
The isolated singlino region (P3) is singlino-dominated and happens for $M_2 < 0$.

For $\tan\beta = 1.03$, there is a small region in the $M_2$-$s$ plane 
with $M_2 > 0$ that satisfies the relic density and LEP chargino mass constraints.
The solution in this region is due to the enhanced $Z \chi^0 \chi^0$ coupling.
For $\tan\beta = 2.0$, there is a large acceptable region with $M_2 <0$.
With $M_2 > 0$ most of the parameter regions that give the relic density are excluded by the LEP 2 chargino search.
For $\tan\beta = 2.5$, there is a large region with $M_2 > 0$ that reproduces the observable relic density and is consistent with the chargino mass limits.

We numerically checked, with suitable parameter values including different $h_s$, that the $S$-model can reproduce the observed relic density for most of the theoretically allowed $m_{\chi^0}$ range without violating the LEP constraints.
However, for a relatively large neutralino mass, {\it i.e.}, $m_{\chi^0} \approx 80$ to $100$ GeV, it becomes hard to satisfy both the relic density and the LEP constraints since those $m_{\chi^0}$ values occur only with $\tan\beta$ close to $1$ and $h_s$ close to its maximum ($0.75$), but then the chargino constraint becomes severe, as can be seen in Figure \ref{fig:contour} (a).

\section{Conclusions}

We have studied the properties of the lightest neutralino in a supersymmetric $U(1)'$ model with a secluded $U(1)'$ breaking sector \cite{Erler:prd.66.015002}.
In this model, $\tan\beta \sim 1$ is required for natural electroweak symmetry 
breaking.
In general, the model allows a lightest neutralino mass up to about $300$ GeV 
(depending on the charge assignments and input mass limits) with the gaugino 
unification assumption $M_{1'} = M_1 = \frac{5}{3} \frac{g_1^2}{g_2^2} M_2$, 
but only up to $m_{\chi^0} \alt 100$ GeV in the limit with 
$M_{1'} \gg M_1 = \frac{5}{3} \frac{g_1^2}{g_2^2} M_2$.
We quantitatively studied the $S$-model in the limit where $M_1{'}$, $s_i$ are much larger than the electroweak scale.
In this limit the $Z \chi^0 \chi^0$ coupling is enhanced compared to the MSSM due to the singlino component.
This allows the $\chi^0$ pair annihilation rate via the $Z$ resonance channel alone to reproduce the observed relic density of the cold dark matter.
In addition, the doublet-singlet mixed nature of the Higgs bosons allows the 
lightest Higgs bosons to have a small mass without violating the LEP 
constraint on SM-like Higgs mass.
The $S$-model explains the relic density over a considerable fraction of the parameter space.

\section*{Acknowledgments}

VB thanks the Aspen Center for Physics for hospitality and CK thanks 
Stan Brodsky, Michael Peskin, and the SLAC Theoretical Physics Group, 
where part of the research was completed, for their hospitality.
This research was supported in part by the U.S. Department of Energy
under Grants 
No.~DE-FG02-95ER40896,
No.~DE-FG02-04ER41305, 
No.~DE-FG02-03ER46040,
and
No.~DOE-EY-76-02-3071, 
and in part by the Wisconsin Alumni Research Foundation.


\end{document}